\documentclass[12pt]{article}

\usepackage{amsmath,amssymb}
\usepackage[english]{babel}
\usepackage[top=20truemm,bottom=20truemm, left= 15truemm,right= 15truemm]{geometry}
\usepackage{graphicx}
\usepackage{times}
\usepackage{bbold}
\usepackage{pifont}
\usepackage{epsfig}
\usepackage{epstopdf}
\usepackage{comment}
\usepackage{bm}
\usepackage{amsmath,amssymb}
\usepackage{authblk}
\usepackage{array}
\usepackage{cite}

\usepackage{color}
\newcommand{\txtr}[1]{\textcolor{red}{#1}}

\def\({\left(}  
\def\){\right)} 
\def\[{\left[}
\def\]{\right]} 
\def\<{\left<} 
\def\>{\right>}

\newcommand{\pd}{{\partial}}

\newcommand{\Tr}{\mathrm{Tr}}

\newcommand{\dd}{{\mathrm{d}}}

\newcommand{\n}{\mathfrak{n}}
\newcommand{\real}{\mathbb{R}} 

\newcommand{\vecS}{\bm{S}}
\newcommand{\w}{\bm{w}}
\newcommand{\W}{\bm{W}}
\newcommand{\Y}{\bm{Y}}
\newcommand{\Z}{\bm{Z}}

\def\ket#1{\left|{#1}\right>}
\def\bra#1{\left<{#1}\right|}

\def\exv#1{\langle{#1}\rangle}

\newcommand{\bhline}[1]{\noalign{\hrule height #1}}

\begin{document}
	\title{\vspace{-2.6cm} Isolated Skyrmions in the $CP^2$ nonlinear $\sigma$-model with a Dzyaloshinskii-Moriya type interaction}
	
	\date{}
	\author[1]{Yutaka Akagi}
	\author[2,3]{Yuki Amari}
	\author[4]{Nobuyuki Sawado}
	\author[2]{Yakov Shnir}

     \affil[1]{\it Department of Physics, Graduate School of Science, The University of Tokyo, Bunkyo, Tokyo 113-0033, Japan}
    \affil[2]{\it BLTP, JINR, Dubna 141980, Moscow Region, Russia}
    \affil[3]{\it Department of Mathematical Physics, Toyama Prefectural University, Kurokawa 5180, Imizu, Toyama, 939-0398, Japan}
    \affil[4]{\it Department of Physics, Tokyo University of Science, Noda, Chiba 278-8510, Japan}
	\maketitle

	\begin{abstract}	
	We study two dimensional soliton solutions in the $CP^2$ nonlinear $\sigma$-model with a Dzyaloshinskii-Moriya type interaction.
	First, we derive such a model as a continuous limit  of the  $SU(3)$ tilted ferromagnetic Heisenberg model on a square lattice. Then, introducing an additional potential term to the derived Hamiltonian, we obtain exact soliton solutions for particular sets of parameters of the model.
	The vacuum of the exact solution can be interpreted as a spin nematic state. For a wider range of coupling constants, we construct numerical solutions, which possess the same type of asymptotic decay as the exact analytical solution, both decaying into a spin nematic state.
		
	\end{abstract} 
	
		\section{Introduction}
	\label{sec:intro}
    \setcounter{equation}{0}
    
    In the 1960s, Skyrme introduced a (3+1)-dimensional $O(4)$ nonlinear (NL)  $\sigma$-model  \cite{Skyrme:1961vq,Skyrme:1962vh} which is now well-known as a prototype of a classical field theory that supports topological solitons (See Ref.~\cite{Manton:2004tk}, for example).  
    Historically, the Skyrme model has been proposed as a low-energy effective theory of atomic nuclei.
    In this framework, the topological charge of the field configuration is identified with the baryon number. 
    
    The Skyrme model, apart from being considered a good candidate for the
low-energy QCD effective theory, has attracted 
much attention in various applications, ranging from string theory and cosmology to condensed matter physics. One of the most interesting developments here is related 
to a planar reduction of the NL$\sigma$-model, the so-called baby Skyrme model \cite{Bogolubskaya:1989ha,Bogolyubskaya:1989fz,Leese:1989gi}. 
This (2+1)-dimensional simplified theory resembles the basic properties of the
original Skyrme model in many aspects.

The baby Skyrme model finds a number of physical realizations in different branches of modern physics. Originally, it was proposed  as a modification of the Heisenberg model \cite{Polyakov:1975yp,Bogolubskaya:1989ha,Bogolyubskaya:1989fz}.  
Then, it was pointed out that Skyrmion configurations naturally
arise in condensed matter systems with intrinsic and induced chirality \cite{bogdanov1989thermodynamically,neubauer2009topological,nagaosa2013topological,leonov2014theory,smalyukh2010three}. These baby Skyrmions, often referred to as magnetic Skyrmions, were  experimentally observed in non-centrosymmetric or chiral magnets \cite{muhlbauer2009skyrmion,yu2010real,heinze2011spontaneous}.
This discovery triggered extensive research 
on Skyrmions in magnetic materials. This direction is a rapidly growing area both theoretically and experimentally \cite{Back_2020}.

    A typical stabilizing mechanism of magnetic skyrmions is the existence of Dzyaloshinskii-Moriya (DM) interaction \cite{dzyaloshinsky1958thermodynamic,moriya1960anisotropic}, which stems from the spin-orbit coupling.
    In fact, the magnetic Skyrmions in chiral magnets can be well described by the continuum effective Hamiltonian   
    \begin{equation}
    H
    = \int \dd^2x
    \[ \frac{J}{2}\(\nabla \bm{m}\) ^2
    + \kappa~\bm{m}\cdot\( \nabla \times \bm{m} \)- Bm^3 + A\left\{|\bm{m}|^2 + \(m^3 \)^2\right\}\],
    \label{Hamiltonian_magnetic_skyrmion}
    \end{equation}
    where $\bm{m}(\bm{r})=\(m^1,m^2,m^3\)$ is a three component unit magnetization vector which corresponds to  the spin expectation value at position $\bm{r}$. 
    The first term in Eq.~\eqref{Hamiltonian_magnetic_skyrmion} is the continuum limit of the Heisenberg exchange interaction, i.e.,
    the kinetic term of the $O(3)$ NL$\sigma$-model, which is often referred to as the Dirichlet term. The second term there is the DM interaction term, the third one is the Zeeman 
    coupling with an external magnetic field $B$, and the last, symmetry breaking term $A\left\{|\bm{m}|^2+\(m^3 \)^2\right\}$ represents the uniaxial anisotropy.

    It is remarkable that in the limiting case $A=\kappa^2/2J, B=0$, the Hamiltonian \eqref{Hamiltonian_magnetic_skyrmion} can be written as the static version of the $SU(2)$ gauged $O(3)$ NL$\sigma$-model \cite{li2011general,Schroers:2019hhe}
    \begin{align}
      H
      =\frac{J}{2} \int \dd^2 x
      \(\pd_k \bm{m} + \bm{A}_k\times\bm{m} \)^2, \qquad k=1,2
      \label{Hamiltonian_A}
    \end{align}
    with a background gauge field 
    $\bm{A}_1=(-\kappa/J, ~0, ~0),~ \bm{A}_2=(0,-\kappa/J,~0)$.
    Though the DM term is usually introduced phenomenologically,
   a mathematical derivation of the Hamiltonian~\eqref{Hamiltonian_A} with arbitrary $\bm{A}_k$ has been developed recently \cite{li2011general}, i.e.; it has been shown that the Hamiltonian can be derived mathematically in a continuum limit of the tilted (quantum) Heisenberg model 
   \begin{equation}
   {\cal H}=-J\sum_{\langle ij \rangle}\({\cal W}_i S^a_i {\cal W}^{-1}_i\)\({\cal W}_j S^a_j {\cal W}^{-1}_j\) ~ ,
   \end{equation}
   where 
   the sum $\langle ij \rangle$ is taken over the nearest-neighbor sites,
   $S^a_i$ denotes the $a$-th component of spin operators at site $i$ and ${\cal W}_i\in SU(2)$. It 
   was reported that the tilting Heisenberg model can be derived from a Hubbard model at half-filling in the presence of spin-orbit coupling \cite{zhu2014spin}.
   Therefore, the background field $\bm{A}_k$ can still be interpreted as an effect of the spin-orbit coupling. 
   
	There are two advantages of utilizing the expression \eqref{Hamiltonian_A} for the theoretical study of baby Skyrmions in the presence of the so-called Lifshitz invariant, an interaction term which is linear in a derivative of an order parameter \cite{sparavigna2009role,yudin2013fundamentals}, like the DM term. 	The first advantage of the form Eq.~\eqref{Hamiltonian_A} is that one can study a NL$\sigma$-model with various form of Lifshitz invariants which are mathematically derived by choice of the background field $\bm{A}_k$, although Lifshitz invariants have, in general, a phenomenological origin corresponding to the crystallographic handedness of a given sample. 
	The second advantage of the model \eqref{Hamiltonian_A} is that it allows us to employ several analytical techniques developed for the gauged NL$\sigma$-model. It has been recently 
	reported in Ref.~\cite{Schroers:2019hhe} that the Hamiltonian \eqref{Hamiltonian_A} with a specific choice of the potential term exactly 
	satisfies the Bogomol'nyi bound, and the corresponding Bogomol'nyi-Prasad-Sommerfield (BPS) equations have exact closed-form solutions \cite{Schroers:2019hhe,Barton-Singer:2018dlh,Ross:2020hsw}.
    
    Geometrically, the planar Skyrmions are very nicely described in terms of the $CP^1$ complex field on the compactified domain space $S^2$ \cite{Leese:1989gi}. Further, there are various generalizations of this model; for example, two-dimensional $CP^2$ Skyrmions have been studied in the pure $CP^2$ NL$\sigma$-model  \cite{Golo:1978de,DAdda:1978vbw,Din:1980jg} and in the Faddeev-Skyrme type model \cite{Ferreira:2010jb,Amari:2015sva}.

    Remarkably, the two dimensional $CP^2$ NL$\sigma$-model can be obtained as a continuum limit of the $SU(3)$ ferromagnetic (FM) Heisenberg model \cite{Ivanov2008pairing,Smerald_2013} on a square lattice defined by the Hamiltonian
	\begin{equation}
	 {\cal H}=-\frac{J}{2}\sum_{\langle ij \rangle}T^m_iT^m_j,
	 \label{Hamiltonian_SU(3)}
	\end{equation}
    where $J$ is a positive constant, and $T^m_i$ ($m=1,...,8$) stand for the $SU(3)$ spin operators of the fundamental representation at site $i$ satisfying the commutation relation 
    \begin{equation}
    	\[T^l_i, T^m_i \]=if_{lmn}T^n_i.
    	\label{com_relation SU(3) operator}
    \end{equation} 
    Here, the structure constants are given by $f_{lmn}=-\frac{i}{2}\Tr\(\lambda_l\[\lambda_m,\lambda_n\]\)$, where $\lambda_m$ are the usual Gell-Mann matrices. 
    
    The $SU(3)$ FM Heisenberg model may play an important role in diverse physical systems ranging from string theory \cite{hernandez20043} to  condensed  matter, or quantum optical three-level systems \cite{greiter2007exact}. It can be derived from a spin-1 bilinear-biquadratic model with a specific choice of coupling constants, so-called FM $SU(3)$ point, see, e.g., Ref.~\cite{Penc2011spin}.
    The $SU(3)$ spin operators can be defined in terms of the $SU(2)$ spin operators $S^a$ ($a=1,2,3$) as
    \begin{equation}
    \(\begin{array}{c}
    T^7\\ 
    T^5\\ 
    T^2
    \end{array}\) 
    =\(\begin{array}{c}
    S^1\\ 
    -S^2\\ 
    S^3
    \end{array}\)
    ,
    \qquad
    \(\begin{array}{c}
    T^3\\ 
    T^8\\ 
    T^1\\ 
    T^4\\ 
    T^6
    \end{array}  \)=-\(\begin{array}{c}
    \(S^1\)^2-\(S^2\)^2\\ 
    \frac{1}{\sqrt{3}}\[\vecS\cdot\vecS-3\(S^3\)^2\]\\ 
    S^1S^2+S^2S^1\\  
    S^3S^1+S^1S^3\\
    S^2S^3+S^3S^2
    \end{array}  \).
    \label{ST}
    \end{equation} 
	Using the $SU(2)$ commutation relation $\[S^a_i, S^b_i \]=i\varepsilon_{abc}S^c_i$ where $\varepsilon_{abc}$ denotes the anti-symmetric tensor, one can check that the operators \eqref{ST} satisfy the $SU(3)$ commutation relation \eqref{com_relation SU(3) operator}.
	
	In the present paper, we study  baby Skyrmion solutions of an extended $CP^2$ NL$\sigma$-model composed of the $CP^2$ Dirichlet term, a DM type interaction term, i.e., the Lifshitz invariant, and a potential term. 
	The Lifshitz invariant, instead of being introduced ad hoc in the continuum Hamiltonian, can be derived in a mathematically well-defined way  via consideration of a continuum limit of the  $SU(3)$ tilted Heisenberg model. Below we will implement this approach in 
	our derivation of the Lifshitz invariant.
    In the extended $CP^2$ NL$\sigma$-model, we derive exact soliton solutions for specific combinations of coupling constants called the BPS point and solvable line. For a broader range of coupling constants, we construct solitons by solving the Euler-Lagrange equation numerically.
	
	The organization of this paper is the following:
	In the next section, we derive an $SU(3)$ gauged $CP^2$ NL$\sigma$-model from the $SU(3)$ tilted Heisenberg model.
	Similar to the $SU(2)$ case described as Eq.~\eqref{Hamiltonian_A}, the term linear in a background field can be viewed as a Lifshitz invariant term. In Sec.~\ref{sec:exact_solution}, we study exact Skyrmionic solutions of the $SU(3)$ gauged $CP^2$ NL$\sigma$-model in the presence of
    a potential term for the BPS point and solvable line using the BPS arguments.
The numerical construction of baby Skyrmion solutions off the solvable line is given in Sec.~\ref{sec:numerical_solution}. 
	Our conclusions are given in 
	Sec.~\ref{sec:conclusion}.

	\section{Gauged $\bm{CP^2}$ NL$\bm{\sigma}$-model from a spin system}
	\label{sec:model_derivation}
	\setcounter{equation}{0}
	
	To find Lifshitz invariant terms relevant for the $CP^2$ NL$\sigma$-model, we begin 
	to derive an $SU(3)$ gauged $CP^2$ NL$\sigma$-model, a generalization of Eq.~\eqref{Hamiltonian_A}, from a spin system on a square lattice.
	By analogy with Eq.~\eqref{Hamiltonian_A}, the Lifshitz invariant, in that case, can be introduced as a term linear in a non-dynamical background gauge potential of the gauged $CP^2$ model.
	
	Following the procedure to obtain a gauged NL$\sigma$-model from a spin system, as discussed in Ref.~\cite{li2011general},
	we consider a generalization of the $SU(3)$ Heisenberg model defined by the Hamiltonian
	\begin{equation}
	{\cal H}=- \frac{J}{2} \sum_{\langle ij \rangle} T^m_i (\hat{U}_{ij})_{mn} T^n_j,
	\label{Hamiltonian_Wilson}
	\end{equation} 
	where $\hat{U}_{ij}$ is a background field which can be recognized as a Wilson line operator along with the link from the point $i$ to the point $j$,  which is an element of the $SU(3)$ group in the adjoint representation.
	As in the $SU(2)$ case \cite{li2011general}, the field $\hat{U}_{ij}$ may describe effects originated from spin (nematic)-orbital coupling, complicated crystalline structure, and so on.
	This Hamiltonian can be viewed as the exchange interaction term for the tilted operator $\tilde{T}^m_i={\cal W}_i T^m_i {\cal W}^{-1}_i$ where ${\cal W}_i\in SU(3)$, because one can write
	${\cal W}_j T^m_j {\cal W}^{-1}_j=(R_j)_{mn}T_j^n$ where $R_j$ is an element of $SU(3)$ in the adjoint representation. Clearly, $\hat{U}_{ij}=R_i^{\rm T} R_j $, where $\rm T$ stands for the transposition.

	Let us now find the classical counterpart of the quantum Hamiltonian \eqref{Hamiltonian_Wilson}.
    It can be defined as an expectation value of 
    Eq.~\eqref{Hamiltonian_Wilson}
    in a state possessing over-completeness, through a path integral representation of the partition function. In order to construct such a state for the spin-1 system, it is convenient to introduce the Cartesian basis
	\begin{equation}
	\ket{x^1}=\frac{i}{\sqrt{2}}\(\ket{+1}-\ket{-1}\),
	\qquad
	\ket{x^2}=\frac{1}{\sqrt{2}}\(\ket{+1}+\ket{-1}\),
	\qquad
	\ket{x^3}=-i\ket{0},
	\label{basis_Cartesian}
	\end{equation}
	where $\ket{m}=\ket{S=1,m}$ ($m=0$, $\pm 1$).
	In terms of the Cartesian basis, an arbitrary spin-1 state at a site $j$ can be expressed as a linear combination $\ket{Z}_j=Z^a(\bm{r}_j)\ket{x^a}_j$
	where ${\bm r}_j$ stands for the position of the site $j$, and  $\bm{Z}=\(Z^1,Z^2,Z^3\)^{\rm T}$ is a complex vector of unit length \cite{Ivanov_2003,Ivanov2008pairing}.  
	Since the state $\ket{Z}_j$ satisfies an over-completeness relation, one can obtain the classical Hamiltonian using the state
	\begin{equation}
	\ket{Z}=\otimes_j\ket{Z}_j=\otimes_j Z^a(\bm{r}_j)\ket{x^a}_j ~.
	\label{state_Z}
	\end{equation}
	Since $\bm{Z}$ is normalized and has the gauge degrees of freedom corresponding to the overall phase factor multiplication, it takes values in $S^5/S^1 \approx CP^2$.	
	In terms of the basis \eqref{basis_Cartesian}, the $SU(3)$ spin operators can be defined as
	\begin{equation}
	T^m=\(\lambda_m \)_{ab}\ket{x^a}\bra{x^b} \qquad m=1,2,\cdots,8,
	\label{Operator_SU(3)}
	\end{equation} 
	where $\lambda_m$ is the $m$-th component of the Gell-Mann matrices.
	One can check that they satisfy the $SU(3)$ commutation relation \eqref{com_relation SU(3) operator}. The expectation values of the $SU(3)$ operators in the state \eqref{state_Z} are given by
	\begin{equation}
	\exv{T^m_j}\equiv n^m(\bm{r}_j)
	=\(\lambda_m \)_{ab} \bar{Z}^a(\bm{r}_j)Z^b(\bm{r}_j),
	\end{equation}
	where 
	$\bar{Z}^a$ denotes the complex conjugation of $Z^a$.
	In the context of QCD, the field $n^m$ is usually termed a color (direction) field \cite{Kondo:2014sta}.
	The color field satisfies the constraints
	\begin{equation}
	n^mn^m = \frac{4}{3},\qquad n^m= \frac{3}{2}d_{mpq}n^pn^q ~,
	\label{constraints_color}
	\end{equation}
	where $d_{mpq}=\frac{1}{4}\Tr\(\lambda_m\left\{\lambda_p,\lambda_q\right\} \)$.
    Consequently, the number of degrees of freedom of the color field reduces to four.
	Note that, combining the constraints~\eqref{constraints_color}, one can get the Casimir identity $d_{mpq}n^mn^pn^q=8/9$.
	
	In terms of the color field, the classical Hamiltonian is given by
	\begin{align}
	H&\equiv\bra{Z}{\cal H}\ket{Z} = - \frac{J}{2} \sum_{\langle ij \rangle} n^l(\bm{r}_i) (\hat{U}_{ij})_{lm} n^m(\bm{r}_j).
	\label{Hamiltonian_classical}	
	\end{align}
	Let us write the position of a site $j$ next to a site $i$ as $\bm{r}_j=\bm{r}_i+a\epsilon \bm{e}_k$ where $\bm{e}_k$ is the unit vector in the $k$-th direction, $\epsilon=\pm 1$, and $a$ stands for the lattice constant.
	For $a\ll1$, the field $\hat{U}_{ij}$ can be approximated by the exponential expansion    
	\begin{equation}
		\hat{U}_{ij}\approx e^{ia\epsilon A^m_k(\bm{r}_i)\hat{l}_m} = {\mathbb 1} + ia\epsilon A_k^m(\bm{r}_i)\hat{l}_m -\frac{a^2}{2} A_k^m(\bm{r}_i)A_k^n(\bm{r}_i)\hat{l}_m \hat{l}_n +{\cal O}(a^3),
	\end{equation}
	where ${\mathbb 1}$ is the unit matrix and $\hat{l}_m$ are the generators of $SU(3)$ in the adjoint representation, i.e., $(\hat{l}_m)_{pq}=if_{mpq}$.
	In addition, since the model \eqref{Hamiltonian_Wilson} is ferromagnetic, it is natural to assume that nearest-neighbor spins are oriented in the almost same direction, which allows us to use the Taylor expansion
	\begin{equation}
	n^m(\bm{r}_j) =
	n^m(\bm{r}_i)+a\epsilon\pd_k n^m(\bm{r}_i)+{\cal O}(a^2).
	\label{Taylor_color}
	\end{equation}
	Replacing the sum over the lattice sites in Eq.~\eqref{Hamiltonian_classical} by the integral $\displaystyle{a^{-2}\int \dd^2x}$, we obtain a continuum Hamiltonian,
	except for a constant term, of the form
	\begin{align}
	H =\frac{J}{8} \int \dd^2x
	\[\Tr \( \pd_k\n\pd_k\n \)-2i \Tr\(A_k\[\n,\pd_k\n \] \) -\Tr\(\[A_k,\n \]^2\) \],
	\label{Hamiltonian_Continuum}
	\end{align}
	where $A_k=A^m_k\lambda_m$ and $\n=n^m\lambda_m$.
	Similar to its $SU(2)$ counterpart expressed as Eq.~\eqref{Hamiltonian_A}, this Hamiltonian can also be written 
	as the static energy of an $SU(3)$ gauged $CP^2$ NL$\sigma$-model
	\begin{equation}
	H=\frac{J}{8}\int \dd^2x \Tr\( D_k \n ~ D_k \n \), 
	\label{Hamiltonian_gaugedCP2}
	\end{equation}
	where $D_k \n= \pd_k \n -i\[A_k,\n \]$ is the $SU(3)$ covariant derivative.
	Since the Hamiltonian is given by the $SU(3)$ covariant derivative, Eq.~\eqref{Hamiltonian_gaugedCP2} is invariant under the $SU(3)$ gauge transformation
	\begin{equation}
	\n\to g\n g^{-1},
	\qquad
	A_k\to gA_kg^{-1}+ig\pd_kg^{-1},
	\label{Gauge_transf}
	\end{equation}
	where $g\in SU(3)$.
	Note that, however, since the Hamiltonian~\eqref{Hamiltonian_gaugedCP2} does not include kinetic terms for the gauge field, like the Yang-Mills term, or the Chern-Simons term, the gauge potential is just a background field, not the dynamical one. We suppose that the gauge field is fixed beforehand by the structure of a sample and give the value by hand, like the $SU(2)$ case. The gauge fixing allows us to recognize the second term in Eq.~\eqref{Hamiltonian_Continuum} as a Lifshitz invariant term.
	
    We would like to 
	emphasize that we do not deal with Eq.~\eqref{Hamiltonian_gaugedCP2} as a gauge theory.  Rather, we deem it the $CP^2$ NL$\sigma$-model with a Lifshitz invariant, and show the existence of the exact and the numerical solutions.
	For the baby Skyrmion solutions we shall obtain, the color field $\n$ approaches to a constant value $\n_\infty$ at spatial infinity so that the physical space $\real^2$ can be topologically compactified to $S^2$. 
	Therefore, they are characterized by the topological degree of the map $\n:\real^2\sim S^2\mapsto CP^2$ given by
	\begin{equation}
	Q = -\frac{i}{32\pi}\int \dd^2x~\varepsilon_{jk}\Tr\(\n \[\pd_j\n,\pd_k \n\]\).
	\label{Topological_charge_n}
	\end{equation}
	Combining with the assumption that the gauge is fixed, it is reasonable to identify this quantity \eqref{Topological_charge_n} with the topological charge in our model\footnote{
	If one extends the model \eqref{Hamiltonian_gaugedCP2} with a dynamical gauge field, the topological charge is defined by the $SU(3)$ gauge invariant quantity which is directly obtained by replacing the partial difference in Eq.~\eqref{Topological_charge_n} with the covariant derivative.}.

	\section{Exact solutions of the $\bm{SU(3)}$ gauged $\bm{CP^2}$ NL$\bm{\sigma}$-model}
	\label{sec:exact_solution}
    \setcounter{equation}{0}
    
	In this section, we derive exact solutions of the model with the Hamiltonian~\eqref{Hamiltonian_gaugedCP2} supplemented by  a potential term.
	We first remark on the validity of the variational problem. 
	As discussed in Refs.~\cite{Schroers:2019hhe,Ross:2020hsw} for the $SU(2)$ case, a surface term, which appears in the process of variation, cannot be ignored if the physical space is non-compact and the gauge potential $A_k$ does not vanish at the spatial infinity like the DM term.  
	This problem can be cured by introducing an appropriate boundary term, like \cite{Schroers:2019hhe}
	\begin{equation}
	H_{\rm Boundary} 
	=\mp4\rho\int \dd^2 x ~\varepsilon_{jk}\pd_j\Tr(\n A_k) \, ,
	\label{Hamiltonian_Boundary}
	\end{equation}
	where $\rho = J/8$.
    Here the gauge potential $A_k$ satisfies
	\begin{equation}
	\[\n_{\infty},A_j \]\pm \frac{i}{2}\varepsilon_{ij}\[\n_\infty,\[\n_\infty, A_k \] \] =0,
	\label{BPS_n_asymptotic}
	\end{equation} 
	where $\n_\infty$ is the asymptotic value of $\n$ at spatial infinity.
	Note that 
	Eq.~\eqref{BPS_n_asymptotic} corresponds to the asymptotic form of the BPS equation, which we shall discuss in the next subsection. Hence, all field configurations we consider in this paper satisfy this equation automatically. 	
	
	Since \eqref{Hamiltonian_Boundary} is a surface term, it does not contribute to the Euler-Lagrange equation, i.e., the classical Heisenberg equation.
	Note that the solutions derived in the following sections satisfy Derrick's scaling relation with the boundary term, which is obtained by keeping the background field $A_k$ intact under the scaling, i. e., $E_1+2E_0=0$ where $E_1$ denotes the energy contribution from the first derivative terms including the boundary term  \eqref{Hamiltonian_Boundary} and $E_0$ from no derivative terms.

	\subsection{BPS solutions}
	\label{subsec:BPS}

	Recently, it has been proved that the $SU(2)$ gauged $CP^1$ NL$\sigma$-model \eqref{Hamiltonian_A} possesses BPS solutions in the presence of a particular potential term \cite{Schroers:2019hhe,Barton-Singer:2018dlh}. 	
	Here, we show that BPS solutions also exist in the $SU(3)$ gauged $CP^2$ model with a special choice of the potential term, which is given by 
	\begin{equation}
	H_{\rm pot}=\pm 4\rho\int \dd^2x   \Tr\(\n F_{12} \),
	\label{pot_BPS}
	\end{equation}
	where $F_{jk}=\pd_j A_k-\pd_k A_j-i\[A_j,A_k \]$.  As we shall see in the next subsection, the potential term can possess a natural physical interpretation for some background gauge field. 
	It follows that the Hamiltonian we study here reads 
	\begin{equation}
	 H=\rho\int \dd^2x  \Tr\(D_k \n ~ D_k \n \) 
	  \pm 4\rho\int \dd^2x   \Tr\(\n F_{12} \) \mp 4\rho\int \dd^2 x ~\varepsilon_{jk}\pd_j\Tr(\n A_k), 
	 \label{Hamiltonian_BPS}
	\end{equation}
    where  the double-sign corresponds to that of Eq.~\eqref{Hamiltonian_Boundary}.
    
	First, let us show that the lower energy bound of Eq.~\eqref{Hamiltonian_BPS} is given by the topological charge~\eqref{Topological_charge_n}.	
	The first term in Eq.~\eqref{Hamiltonian_BPS} can be written as 
	\begin{align}
	\rho\int \dd^2x ~\Tr\(D_k \n ~ D_k \n \)   
	&=\frac{\rho}{2}\int \dd^2x \[ \Tr\(D_k \n ~ D_k \n \)+\(\frac{i}{2}\)^2\Tr\(\[\n,D_k \n\]^2\) \]
	\notag\\
	&=\frac{\rho}{2}\int \dd^2x  \Tr\(D_j \n \pm \frac{i}{2} \varepsilon_{jk}\[\n,D_k \n\]\)^2 
	\pm\frac{i\rho}{2} \int \dd^2x  ~\varepsilon_{jk}\Tr\(\n \[D_j\n,D_k \n\]\)
	\notag\\
	 &\geq
	 \pm\frac{i\rho}{2} \int \dd^2x  ~\varepsilon_{jk}\Tr\(\n \[D_j\n,D_k \n\]\) .
	\end{align}
	It follows that the equality is satisfied if
	\begin{equation}
	D_j \n \pm \frac{i}{2} \varepsilon_{jk}\[\n,D_k \n\]=0,
	\label{BPS_n}
	\end{equation} 
	which reduces to  Eq.~\eqref{BPS_n_asymptotic} at the spatial infinity. Therefore, one 
	obtains the lower bound of the form 
	\begin{align}
	H&\geq \pm\frac{\rho}{2} \int \dd^2x \[ i\varepsilon_{jk}\Tr\(\n \[D_j\n,D_k \n\]\) +8\Tr\(\n F_{12} \)-8\varepsilon_{jk}\pd_j\Tr\(\n A_k \)\]
	\notag\\
	&= \pm \frac{i\rho}{2}\int \dd^2x~ \varepsilon_{jk}
	 \Tr\(\n \[\pd_j\n,\pd_k \n\]\)  
	\notag\\
	&= \mp 16\pi \rho ~Q ,
	\label{Bound}
	\end{align}
	where the corresponding BPS equation is given by Eq.~\eqref{BPS_n}.
	Note that, unlike the energy bound of the  $CP^N$ self-dual solutions \cite{Polyakov:1975yp,DAdda:1978vbw}, the energy bound \eqref{Bound} can be negative, and it is not proportional to the absolute value of the topological charge. 
	
	As is often the case in two-dimensional BPS equations \cite{Polyakov:1975yp,Schroers:2019hhe}, solutions can be best described in terms of the complex coordinates $z_\pm=x^1\pm ix^2$. 
	Further, we make use of the associated differential operator and background field defined as $\pd_\pm=\frac{1}{2}\(\pd_1\mp i\pd_2 \)$ and $A_\pm=\frac{1}{2}(A_1\mp iA_2)$.
	Then, the BPS equation \eqref{BPS_n} can be written as
	\begin{equation}
	D_\pm \n - \frac{1}{2} \[\n,D_\pm \n\]=0.
	\label{BPS_pm}
	\end{equation} 
	Similar to the $SU(2)$ case \cite{Schroers:2019hhe}, Eq.~ \eqref{BPS_pm} with a plus sign can be solved if the background field has the form 
	\begin{equation}
	A_+=ig^{-1}\pd_+ g, 
	\label{Gauge_Holomporphic}
	\end{equation} 
	where $g\in SL(3,\mathbb{C})$. Note that Eq.~\eqref{Gauge_Holomporphic} is not necessarily a pure gauge.
	Similarly, 
	Eq.~\eqref{BPS_pm} with the minus sign on the right-hand side can be solved if $A_-=ig^{-1}\pd_- g$. 
	For the background field \eqref{Gauge_Holomporphic}, one finds that the BPS equation \eqref{BPS_pm} is equivalent to 
	\begin{equation}
	\pd_+ \tilde{\n} - \frac{1}{2} \[\tilde{\n},\pd_+ \tilde{\n} \]=0,~~
	\tilde{\n}=g\n g^{-1},
	\label{BPS_pd_+}
	\end{equation}
because, under the $SL(3,\mathbb{C})$ gauge transformation, the fields are changed as $\n \to \tilde{\n} = g\n g^{-1}$ and $A_+\to \tilde{A}_\pm=gA_+g^{-1}+ig\pd_\pm g^{-1}=0$. In the following, we      only consider Eq.~\eqref{Gauge_Holomporphic} to simplify our discussion.
	
	In order to solve the equation \eqref{BPS_pd_+}, we introduce a tractable parameterization of the color field
	\begin{equation}
	\n=-\frac{2}{\sqrt{3}}U\lambda_8 U^\dagger,
	\end{equation} 
	with $U=\(\bm{Y}_1,\bm{Y}_2,\bm{Z} \) \in SU(3)$, where $\bm{Z}$ is the continuum counter part of the vector $\bm{Z}$ in Eq.~\eqref{state_Z} and $\bm{Y}_1,\bm{Y}_2$ are vectors forming an orthonormal basis for $\mathbb{C}^3$ with $\bm{Z}$. 
	Up to the gauge degrees of freedom, the components $\bm{Y}_i$ can be written as
	\begin{equation}
	\bm{Y}_1=\frac{\(-\bar{Z}^3,0,\bar{Z}^1 \)^{\rm T}}{\sqrt{1-|Z^2|^2}}, 
	\qquad
	\bm{Y}_2=\frac{\(-\bar{Z}^2Z^1,1-|Z^2|^2, -\bar{Z}^2Z^3 \)^{\rm T}}{\sqrt{1-|Z^2|^2}}.
	\end{equation} 
	Therefore, the vector $\bm{Z}$ fully defines the color field $\n$.
	Accordingly, we can write
	\begin{equation}
	\tilde{\n}=-\frac{2}{\sqrt{3}}W\lambda_8 W^{-1},
	\label{parametrization_ntilde}
	\end{equation}
	with $W=gU=\(\W_1, \W_2, \W_3 \) \in SL(3,\mathbb{C})$.
	It follows that the field $\bm{Z}$, which is the fundamental field of the model, is given by $\bm{Z}=g^{-1}\W_3$.
	Substituting the field \eqref{parametrization_ntilde} into the equation \eqref{BPS_pd_+}, one finds that Eq.~\eqref{BPS_pd_+} reduces to the coupled equation
	\begin{equation}
	\left\{
	\begin{aligned}
	\W_1^{-1}\pd_+ \W_3=0 
	\\
	\W_2^{-1}\pd_+ \W_3=0 
	\end{aligned}
	\right. ,
	\label{BPS_W}
	\end{equation}
	where $\W_l^{-1}=\Y_l^\dagger g^{-1}$ $(l=1,2)$.
	Since the three vectors $\{ \Y_1,\Y_2,\Z \}$ form an orthonormal basis, Eq. \eqref{BPS_W} implies
	$ \pd_+\bm{W}_3 =\beta \bm{W}_3$
	where the function $\beta$ is given by $
	\beta =\beta \W^{-1}_3\W_3=\W^{-1}_3\pd_+\bm{W}_3.
	$
	Therefore, the equation \eqref{BPS_pd_+} is solved by any configuration satisfying
	\begin{equation}
	{\cal D}_+\W_3 = 0, 
	\label{BPS_DW}
	\end{equation}
	where ${\cal D}_+ {\bm \Phi} = \pd_+ {\bm \Phi} -({\bm \Phi}^{-1}\pd_+{\bm \Phi}) {\bm \Phi}$ for arbitrary non-zero vector $\bm \Phi$.
	Moreover, we write
	\begin{equation}
	\bm{W}_3= \sqrt{|\bm{W}_3|^2} ~ \w,
	\end{equation}
	where $\w$ is a three component unit vector, i.e. $|\w|^2=\w^\dagger \w=1$.
	Then, Eq.~\eqref{BPS_DW} can be reduced to
	\begin{equation}
	{\cal D}_+\w\equiv \pd_\mu\w-\(\w^\dagger \pd_\mu\w \)\w=0,
	\end{equation}
	which is the very BPS equation of the standard $CP^2$ NL$\sigma$-model. 
	Thus, a general solution of 
	Eq.~\eqref{BPS_DW}, up to the gauge degrees of freedom, is given by
	\begin{equation}
	\w=\frac{{\bm P}}{|{\bm P}|}, \qquad 
	{\bm P}=\(P_1(z_-),P_2(z_-),P_3(z_-) \)^{\rm T},
	\end{equation} 
	where ${\bm P}$ has no overall factor, and $P_a$ is a polynomial in $z_-$.
	Therefore, we finally obtain the solution for the $\bm{Z}$ field
	\begin{equation}
	\Z=g^{-1}\W_3
	=\chi g^{-1}\w
	=\chi g^{-1}{\bm P},
	\label{sol_BPS}
	\end{equation}
	where $\chi$ is a normalization factor.

	\subsection{Properties of the BPS solutions}
	\label{subsec:properties_BPS}

	As the BPS bound~\eqref{Bound} indicates, the lowest energy solution among Eq.~\eqref{sol_BPS} with a given background function $g$ possesses the highest topological charge. In terms of the explicit calculation of the topological charge, we discuss the conditions for the lowest energy solutions. 
	
	The topological charge \eqref{Topological_charge_n} can be written in terms of $\Z$ as 
	\begin{equation}
	Q=-\frac{i}{2\pi}\int \dd^2 x ~\varepsilon^{ij}\({\cal D}_i \Z\)^\dagger {\cal D}_j \Z.
	\label{TopCh_Z}
	\end{equation}
	We employ the constant background gauge field  $A_+$ for simplicity.
	Then, the matrix $g$ in Eq.~\eqref{Gauge_Holomporphic} becomes 
	\begin{equation}
		g=\exp\(-iA_+ z_+ \)\, ,
	\end{equation}
	so that the components of $g^{-1}$ are given by power series in $z_+$. 
	It allows us to write Eq.~\eqref{TopCh_Z} as a line integral along the circle at spatial infinity
	\begin{equation}
	Q = \frac{1}{2\pi}\int_{S^1_\infty} C,
	\label{TopCh_C}
	\end{equation}
	with $C=-i\Z^\dagger \dd \Z$ \cite{DAdda:1978vbw,Zakrzewski:1989na}, since the one-form $C$ becomes  globally well-defined.
	To evaluate the integral in Eq.~\eqref{TopCh_C}, we write explicitly
	\begin{equation}
	\Z=\frac{\chi}{\sqrt{|P_1|^2+|P_2|^2 + |P_3|^2}} \sum_a \(\begin{array}{c}
	g^{-1}_{1a}(z_+)P_a(z_-)\\
	g^{-1}_{2a}(z_+)P_a(z_-) \\ 
	g^{-1}_{3a}(z_+)P_a(z_-)
	\end{array} \),  
	\end{equation}
	where $g^{-1}_{ab}$ is the $(a,b)$ component of the inverse matrix $g^{-1}$.
	
	Let $N_a$ ($K_{ab}$) be the highest power in $P_a$ ($g^{-1}_{ab}$). 
	Note that though $g^{-1}_{ab}$ are formally represented as power series in $z_+$, the integers $K_{ba}$ are not always infinite; especially, if a positive integer power of $A_+$ is zero, all of $K_{ba}$ become finite because $g^{-1}$ reduces to a polynomial of finite degree in $z_+$.
	Using the plane polar coordinates $\{r,\theta\}$, one can write  
	$g^{-1}_{ba}(z_+)P_a(z_-)\sim r^{N_a+K_{ba}}\exp[-i(N_a-K_{ba})\theta]$ at the spatial boundary and find that only 
	the components of the highest power in $r$ 
	contribute to the integral~\eqref{TopCh_C}. 	
	Since we are interested in constructing  topological solitons, we consider the case when the physical space $\real^2$ can be compactified to the sphere $S^2$, i.e., the field $\bm{Z}$ 
	takes some fixed value on the spatial boundary.
	Such a  compactification is possible if there is only one pair $\{N_a,K_{ba}\}$ giving the largest sum $N_a+K_{ba}$ or any pairs $\{N_a,K_{ba}\}$, sharing the largest sum, have the same value of the difference.
	For such configurations, the topological charge is given by
	\begin{equation}
	Q = -N_a+K_{ba}, 
	\label{Q_DM}
	\end{equation}  
	where the combination $\{N_a,K_{ba}\}$ yields the largest sum among any pairs $\{N_c,K_{dc}\}$. 
	This equation~\eqref{Q_DM} indicates that the highest topological charge configuration is given by the choice $N_a=0$ for a particular value of $a$ which gives the biggest $K_{ba}$.
		
	\begin{figure}[t!]
		\centering
		\includegraphics[width=8cm,clip]{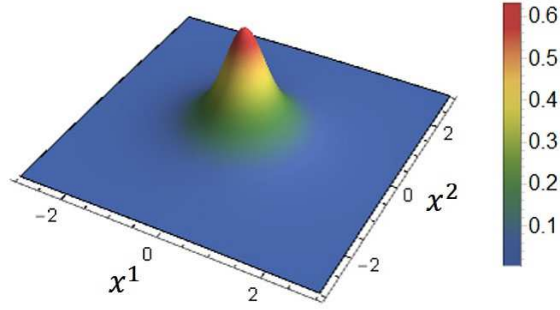}
		\caption{Topological charge density of the axial symmetric solution \eqref{solution_nematic_simplest} with $\kappa=1$. \label{Fig_qdens} }
	\end{figure}
	
\begin{figure}[t!]
		\centering
		\includegraphics[width=18cm,clip]{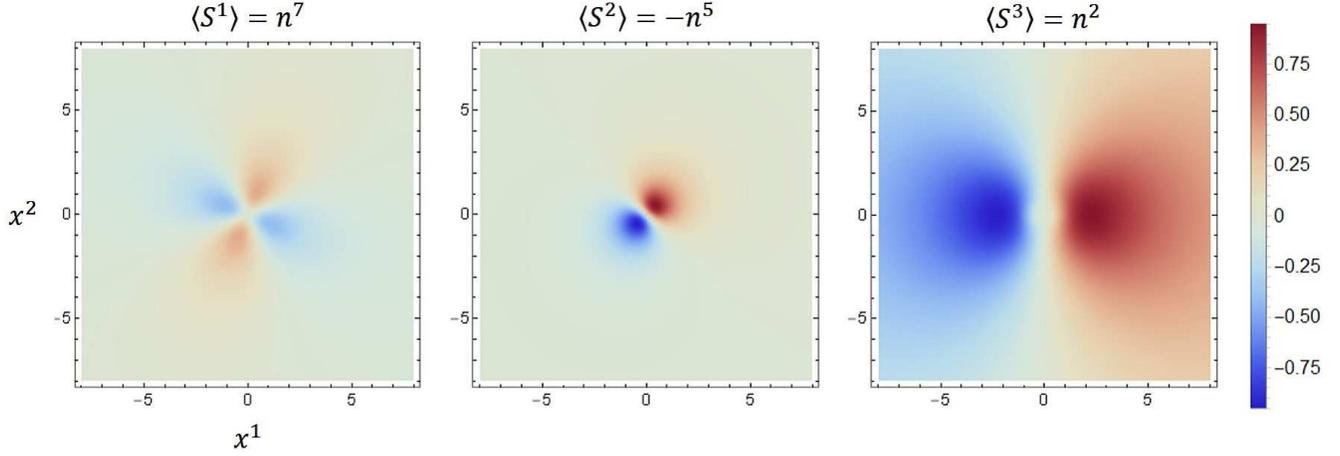}
		\caption{The expectation values $\exv{S^a}$ for the solution \eqref{solution_nematic_simplest} with $\kappa=1$.  \label{Fig_expS} }
	\end{figure}

\begin{figure}[t!]
		\centering
		\includegraphics[width=18cm,clip]{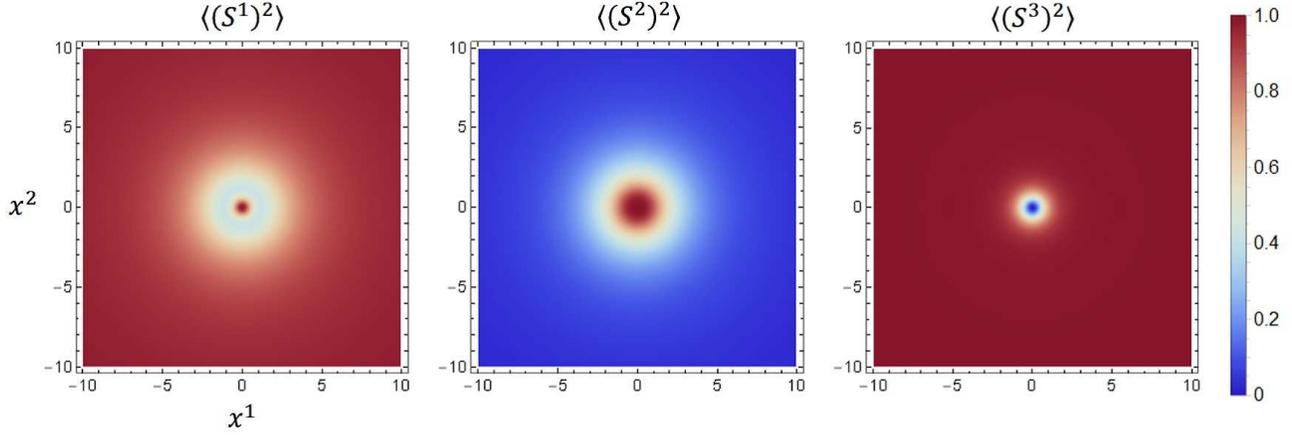}
		\caption{The expectation values $\exv{\(S^a\)^2}$ for the solution \eqref{solution_nematic_simplest} with $\kappa=1$.   \label{Fig_expS2} }
	\end{figure}

	We are looking for the lowest energy solutions with an explicit background field. As a particular example, let us consider 
	\begin{equation}
	A_1= \kappa \(\lambda_1 + \lambda_4 + \lambda_5\), \qquad 
	A_2 = \kappa\(\lambda_2 + \lambda_4 - \lambda_5\),
	\label{gaugepot_nematic}
	\end{equation}
	where $\kappa$ is a constant. Clearly, this choice  yields the potential term
	\begin{equation}
	V=4\Tr\(\n F_{12} \)=-16\sqrt{3}\kappa^2n^8 =16\kappa^2\(2-3\exv{(S^3)^2}\),
	\end{equation}
	which can be interpreted as an easy-axis anisotropy, or quadratic Zeeman term, which naturally appears in condensed matter physics. 
	In this case, the solution \eqref{sol_BPS} can be written as
	\begin{equation}
	\bm{Z}=\frac{\chi}{\sqrt{\Delta}}\(\begin{array}{c}
	P_1(z_-) + \sqrt{2}\kappa z_+ e^{\frac{\pi i}{4}}P_3(z_-)  \\ 
	P_2(z_-) + i\kappa z_+ P_1(z_-) + \frac{\kappa^2 z_+^2}{\sqrt{2}}e^{\frac{3\pi i}{4}} P_3(z_-)\\ 
	P_3(z_-)
	\end{array}  \).
	\end{equation}
	Therefore, the solution with the highest topological charge is given by $P_1=\alpha_1$, $P_2=\alpha_2 z_-+\alpha_3$ with $\alpha_i \in\mathbb{C}$, and  $P_3$ being a nonzero constant.
	Choosing $P_1=P_2=0$, one can 
	obtain the  axially-symmetric solution 
	\begin{equation}
	\bm{Z}=\frac{1}{\sqrt{\Delta}}\(\begin{array}{c}
	\sqrt{2}\kappa z_+ e^{\frac{\pi i}{4}}  \\ 
	\frac{\kappa^2 z_+^2}{\sqrt{2}}e^{\frac{3\pi i}{4}} \\ 
	1
	\end{array}  \), \qquad \Delta = 1 + 2\kappa^2 z_+z_- + \frac{\kappa^4}{2}z_+^2z_-^2,
	\label{solution_nematic_simplest}
	\end{equation}
	which possesses the topological charge $Q=2$. Note that this configuration also satisfies the BPS equation of the pure $CP^2$ NL$\sigma$-model \cite{Golo:1978de,DAdda:1978vbw,Ivanov2008pairing}. 
	Figure \ref{Fig_qdens} shows the distribution of the topological charge \eqref{TopCh_Z} of this solution \eqref{solution_nematic_simplest} with $\kappa=1$.
	We find that the topological charge density has a single peak, although higher charge topological solitons with axial symmetry are likely to possess a volcano structure, see e.g., Ref.~\cite{Piette:1994ug}.
	These highest charge solutions give the asymptotic values at spatial infinity of the color field  
	\begin{equation}
	\(n_\infty^1, n_\infty^2, n_\infty^3, n_\infty^4, n_\infty^5, n_\infty^6, n_\infty^7, n_\infty^8 \) = (0,0,-1,0,0,0,0,1/\sqrt{3}) ~.
	\label{vacuum}
	\end{equation}
	It indicates that $\n$ takes the vacuum value in the Cartan subalgebra of $SU(3)$.
	Hence, the vacuum of the model corresponds to a spin nematic, i.e., $\exv{S^1}=\exv{S^2}=\exv{S^3}=0$ and $\exv{\(S^2\)^2}=0, \exv{\(S^1\)^2}=\exv{\(S^3\)^2}=1$.
	Unlike the pure $CP^2$ model, there is no degeneracy between the spin nematic state and ferromagnetic state in our model because the $SU(3)$ global symmetry is broken. 
    As shown in Fig.~\ref{Fig_expS}, the spin nematic state is partially broken around the soliton because the expectation values $\exv{S^a}$ become finite.
	Fig.~\ref{Fig_expS2} shows that $\exv{\(S^a\)^2}$ of the solution \eqref{solution_nematic_simplest} are axially symmetric, although the expectation values $\exv{S^a}$ have angular dependence.
	
	
	\subsection{Exact solutions off the BPS point}
	\label{sec:solvable_line}

	Note that the Hamiltonian \eqref{Hamiltonian_magnetic_skyrmion} with $B=2A$ admits closed-form analytical solutions  \cite{doring2017compactness}. Further, the $CP^1$ BPS truncation   
	corresponds to the restricted choice of the parameters, $B=2A=\kappa^2$. 
	The relation $B=2A$ is referred to as  the solvable line, whereas the restriction  $B=2A=\kappa^2$ is called the BPS point \cite{Ross:2020hsw}.
	Here we show that similar restrictions occur in our model.
	For this purpose, we consider the generalized Hamiltonian
	\begin{equation}
	H = H_{\rm D} + H_{\rm L} + H_{\rm Boundary} + \nu^2 H_{\rm ani} + \mu^2H_{\rm pot},
	\label{Hamiltonian_generaln}
	\end{equation}
	where $\nu$ and $\mu$ are real coupling constants. Here, $H_{\rm D}$  
	indicates the $CP^2$ Dirichlet term, i.e., the first term in the r.h.s of Eq.~\eqref{Hamiltonian_Continuum}, and $H_{\rm L}$ does the Lifshitz invariant term which is the second term of that. Explicitly, these and other terms read
	\begin{align}
	& H_{\rm D} = \rho\int \dd^2x \Tr\(\pd_k\n ~\pd_k \n \),
	\label{Hamiltonian_Dirichlet}
	\\
	&H_{\rm L} = -2i \rho \int \dd^2x \Tr\(A_k\[\n,\pd_k\n \] \),
	\label{Hamiltonian_Lifshitz}
	\\
	&H_{\rm ani} = - \rho \int \dd^2x \[ \Tr\(\[A_k,\n \]^2 \) - \Tr\(\[A_k,\n_\infty \]^2 \) \],
	\label{Hamiltonian_ani}
	\\
	&H_{\rm pot} = 4\rho\int \dd^2x  \[ \Tr\(\n F_{12} \) - \Tr\(\n_\infty F_{12} \)\],
	\label{Hamiltonian_potential} \,
	\end{align}
    where $A_k$ is a constant background field, as before. Finally, the boundary term $H_{\rm Boundary}$ is defined by Eq.~\eqref{Hamiltonian_Boundary} with the negative sign in the r.h.s., the same as before. 
	Note that we also introduced constant terms in Eqs.~\eqref{Hamiltonian_ani} and \eqref{Hamiltonian_potential} in order to  guarantee the finiteness  of the total energy. Clearly, the Hamiltonian \eqref{Hamiltonian_generaln} is reduced to Eq.~\eqref{Hamiltonian_BPS} as we set  $\nu^2=\mu^2=1$.
	
	The existence of exact solutions of the Hamiltonian~\eqref{Hamiltonian_generaln} with $\nu^2=\mu^2$ can be easily shown if we rescale the space coordinates as $\vec{x}\to r_0\vec{x}$, where $r_0$ is a positive constant, while the background gauge field $A_k$ remains intact.	
	By rescaling, the Hamiltonian \eqref{Hamiltonian_generaln}  becomes 
	\begin{equation}
	H = H_{\rm D} + r_0\(H_{\rm L} + H_{\rm Boundary}\) + r_0^2 \(\nu^2H_{\rm ani} + \mu^2H_{\rm pot} \).
	\end{equation}
	Setting $\nu^2=\mu^2$ and  choosing the scale parameter $r_0=\nu^{-2}$, one gets 
	\begin{equation}
	H^{r_0=\nu^{-2}}_{\nu^2=\mu^2} = H_{\rm D} + \nu^{-2}\(H_{\rm L} + H_{\rm Boundary} + H_{\rm ani} + H_{\rm pot}\).
	\end{equation}
	
	Notice that since the solutions \eqref{sol_BPS} with $P_i$ being arbitrary constants  are holomorphic maps from $S^2$ to $CP^2$, they  satisfy not only the variational equations $\delta H_{\nu^2=\mu^2=1}=0$ but also the equations $\delta H_{\rm D} =0$, where $\delta$ denotes the variation with respect to $\n$ with preserving the constraint \eqref{constraints_color}.
	Therefore,  the solutions also satisfy the equations $\delta H^{r_0=\nu^{-2}}_{\nu^2=\mu^2}=0$. 
	This implies that, in the limit $\mu^2=\nu^2$,  the Hamiltonian \eqref{Hamiltonian_generaln} supports a family of exact solutions of the form 
	\begin{equation}
	\bm{Z}(\nu^2) = \exp\[i\nu^2A_+z_+\]\bm{c}\, ,
	\label{sol_nu}
	\end{equation}
	where $\bm{c}$ is a three-component complex unit vector.
	
	Since the solution \eqref{sol_nu} is a BPS solution of the pure $CP^2$ model with the positive topological charge $Q$, one gets $H_{\rm D}[\bm{Z}(\nu^2)]=16\pi \rho Q$. 
	In addition, the lower bound at the BPS point \eqref{Bound} indicates that $H_{\nu^2=\mu^2=1}[\bm{Z}(\nu^2=1)]= -16\pi\rho Q$.
	Combining these bounds, we find that the total energy of the solution \eqref{sol_nu} is given by
	\begin{align}
	H_{\nu^2=\mu^2}[\bm{Z}(\nu^2)] =16\pi\rho\(1-\frac{2}{\nu^{2}}\)Q.
	\end{align} 
    Since the energy becomes negative if $\nu^2 < 2$, we can expect that for small values of the coupling $\nu^2$, the homogeneous vacuum state becomes unstable, 
    and then
    separated 2D Skyrmions (or a Skyrmion lattice) emerges as a ground state.  
	
	\section{Numerical solutions}
	\label{sec:numerical_solution}
	\setcounter{equation}{0}
	
	\subsection{Axial symmetric solutions}

	In this section, we study baby Skyrmion solutions of the Hamiltonian \eqref{Hamiltonian_generaln} with various combinations of the coupling constants.
	Apart from the solvable line, no exact solutions could find analytically, and then we have to solve the equations numerically.
	Here, we restrict ourselves to  the case of the background field given by Eq.~\eqref{gaugepot_nematic}. 

	For the background field \eqref{gaugepot_nematic}, by analogy with the case of the single $CP^1$  magnetic Skyrmion solution, we can look for a configuration described by the axially symmetric ansatz
	\begin{equation}
	\bm{Z}=\( \sin F(r) \cos G(r) e^{i\Phi_1(\theta)}, \sin F(r) \sin G(r) e^{i\Phi_2(\theta)}, \cos F(r)  \),
	\label{Ansatz_axial_sym}
	\end{equation}
	where $F$ and $G$ ($\Phi_1$ and $\Phi_2$) are real functions of the plane polar coordinates $r$ ($\theta$).
	
	The exact solution on the solvable line $\nu^2=\mu^2$ with axial symmetry can be written in terms of the ansatz with the functions
	\begin{equation}
	F = \tan^{-1}\sqrt{2\nu^4\kappa^2 r^2 + \frac{\nu^8\kappa^4r^4}{2} },
	\qquad
	G = \tan^{-1}\(\frac{\nu^2\kappa r}{2} \),
	\qquad
	\Phi_1= \theta+\frac{\pi}{4},
	\qquad
	\Phi_2= 2\theta+\frac{3\pi}{4}.
	\label{sol_nonBPS_holomorphic}
	\end{equation}
	Further, the solution~\eqref{solution_nematic_simplest} is given by Eq.~\eqref{sol_nonBPS_holomorphic} with $\nu^2=1$.	
	This configuration is a useful reference point in the configuration space as we discuss below some properties of numerical solutions 
	in the extended model~\eqref{Hamiltonian_generaln}.
	
	For our numerical study, it is convenient to introduce the energy unit $8\rho$ and the length unit $\kappa^{-1}$, in order to scale 
	the coupling constants.	
	Then, the rescaled components of the Hamiltonian with the ansatz \eqref{Ansatz_axial_sym} become 
	\begin{align}
	& H_{\rm D} =  \int \dd^2x \[ F'^2 + \sin^2 F G'^2 + \right. 
	\notag\\ 
	&\left. \qquad\qquad\quad
	+ \frac{\sin^2F}{r^2}\left\{\dot{\Phi}_1^2 \cos^2G + \dot{\Phi}_2^2 \sin^2G\right\} - 
	\frac{\sin^4F}{r^2}\(\dot{\Phi}_1 \cos^2G + \dot{\Phi}_2 \sin^2G\)^2 \],
	\\
	& H_{\rm L} = -2\int\frac{\dd^2x}{r}
	\Bigg[\sqrt{2}\cos \(\theta +\frac{\pi }{4}-\Phi_1\) 
	\left\{ r\(\cos G F'- \sin2 F \sin G\frac{G'}{2} \)
	+\sin2F \cos G \frac{\dot{\Phi}_1}{2} 
	\right.
	\notag\\
	&\qquad\qquad\qquad\qquad\qquad\qquad\qquad\qquad\qquad
	-\sin2F \sin ^2F\cos G \(\cos^2G \dot{\Phi}_1+\sin ^2G \dot{\Phi}_2\)
	\Big\}
	\notag\\
	&\qquad\qquad\qquad\qquad\quad
	-\sin \(\theta+\Phi_1-\Phi_2\) \left\{r \sin^2F	G'+\frac{1}{2}\sin^2F\sin 2G \(\dot{\Phi}_1+\dot{\Phi}_2\) 
	\right.
	\notag\\
	& \qquad\qquad\qquad\qquad\qquad\qquad\qquad\qquad\quad\qquad\qquad
	-\sin^4F\sin 2G\left(\cos ^2G \dot{\Phi}_1+\sin^2G \dot{\Phi}_2\)\Big\} \Bigg],
	\\
	& H_{\rm ani} = \frac{1}{2}\int\dd^2x\Bigg[
	16\sin^2 F \cos^2G\left\{\cos^2F -\frac{1}{\sqrt{2}}\cos\(2\Phi_1-\Phi_2+\frac{\pi}{4}\)\sin 2F\sin G + \sin^2 F \sin G^2\right\}
	\notag\\
	& \qquad\qquad\qquad\qquad
	+\sin^22F(1 + 2\sin^2 G) + 8(\cos^2F - \cos^2G\sin^2F)^2 + 4\cos^22G\sin^4F-4 \Bigg],
	\\
	& H_{\rm pot}= 2\int \dd^2x \( 1- \sqrt{3}n^8 \) = 6\int \dd^2x \cos^2F,
	\end{align}
	where the prime $'$ and the dot $\dot{}$ stands for the derivatives with respect to the radial coordinate $r$ and angular coordinate $\theta$, respectively.
	The system of corresponding Euler-Lagrange equations for $\Phi_i$ can be solved algebraically for an arbitrary set of the coupling constants, and the solutions are 
	\begin{equation}
	\Phi_1=\theta+\frac{\pi}{4},\qquad
	\Phi_2=2\theta+\frac{3\pi}{4}+m\pi \, ,
	\label{Phase_factor}
	\end{equation}
	where $m$ is an integer. Without loss of generality, we choose $m=0$ by transferring the corresponding multiple windings of the phase $\Phi_2$ to the sign of the profile function $G$.
	Then, the system of the Euler-Lagrange equations for the profile functions with the phase factor \eqref{Phase_factor}  reads
	\begin{equation}
	\begin{split}
	&\frac{\delta H_{\rm D}}{\delta F} 
	+ \frac{\delta H_{\rm L}}{\delta F} 
	+ \nu^2\frac{\delta H_{\rm ani} }{\delta F} 
	+ \mu^2 \frac{\delta H_{\rm pot} }{\delta F}=0,
	\\
	&\frac{\delta H_{\rm D}}{\delta G}
	+\frac{\delta H_{\rm L}}{\delta G} 
	+ \nu^2\frac{\delta H_{\rm ani} }{\delta G} 
	+ \mu^2 \frac{\delta H_{\rm pot} }{\delta G} =0,
	\end{split}
	\label{equation_profile}
	\end{equation}
	with 
	\begin{align}
	&\frac{\delta H_{\rm D}}{\delta F} =
	\[ 2rF''+2F'-\sin 2F\left\{ rG'^2+\frac{1+3\sin^2G}{r} - \frac{2\sin^2F}{r}\(1+\sin^2G \)^2 \right\}   \], 
	\\
	&\frac{\delta H_{\rm L}}{\delta F} =
	-2\[ 2\sqrt{2}\sin^2F\{-r\sin G G'+\cos G +\cos G\(1+\sin^2 G\)\(4\cos^2F -1 \) \} \right. 
	\notag\\
	&\left. \qquad\qquad\qquad\qquad
	-r\sin 2F G'-\frac{3}{2}\sin 2F \sin 2G +4\cos F \sin^3F \sin 2G \(1+\sin^2G \)  \],
	\\
	&\frac{\delta H_{\rm ani} }{\delta F} =2r\[ 
	4\sqrt{2}\sin G\cos^2 G \sin^2 F\(3-4\sin^2F \) 
	-4\cos F\sin^3 F \cos^2 2G
	\right.
	\notag\\
	&\left.  \qquad\qquad\qquad
	+4\sin 2F\left\{ \cos^2F - \sin^2F\cos^2G\(1+\sin^2G\) \right\}
	- \sin 2F\cos 2F (1+ 2\sin^2 G)
	\],
	\\
	&\frac{\delta H_{\rm pot} }{\delta F} = 6r\sin 2F,
	\\
	&\frac{\delta H_{\rm D}}{\delta G} = 
	\[ 2r\sin F^2 G''+2r\sin 2F F'G' + 2\sin^2FG' 
	- \frac{\sin^2F\sin 2G}{r}\left\{ 3 - 2\sin^2F\(1+\sin^2G \) \right\} \],
	\\
	&\frac{\delta H_{\rm L}}{\delta G} = 
	-2\[ \sqrt{2}\sin^2F\sin G \left\{2rF'+\sin 2F \(1-3\sin^2G\)  \right\} 
	\right.
	\notag\\
	&\left. \qquad\qquad\qquad\qquad
	+r\sin 2FF'+ \sin^2F(1-3\cos 2G) + \sin^4F\(1 +3\cos 2G - 2 \cos^22G \) \],	
	\\
	& \frac{\delta H_{\rm ani} }{\delta G} = 
	r\[8\sqrt{2}\cos F\sin^3 F \cos G\(1 - 3\sin^2G\) 
	+ 16\sin^4F\cos^3G\sin G - \sin^22F\sin 2G \],
	\\
	& \frac{\delta H_{\rm pot} }{\delta G} =0. 
	\end{align}
    We solve the equations for $\nu^2\neq \mu^2$ numerically with the boundary condition
    \begin{equation}
    	F(0)=G(0)=0,\qquad \lim_{r\to\infty} F(r)=\lim_{r\to\infty} G(r)= \pi/2,
    	\label{BC}
    \end{equation}
    which the exact solution \eqref{sol_nonBPS_holomorphic} satisfies.
    This vacuum corresponds to the spin nematic state \eqref{vacuum}.

	\begin{figure}[t!]
		\centering
		\includegraphics[width=16cm,clip]{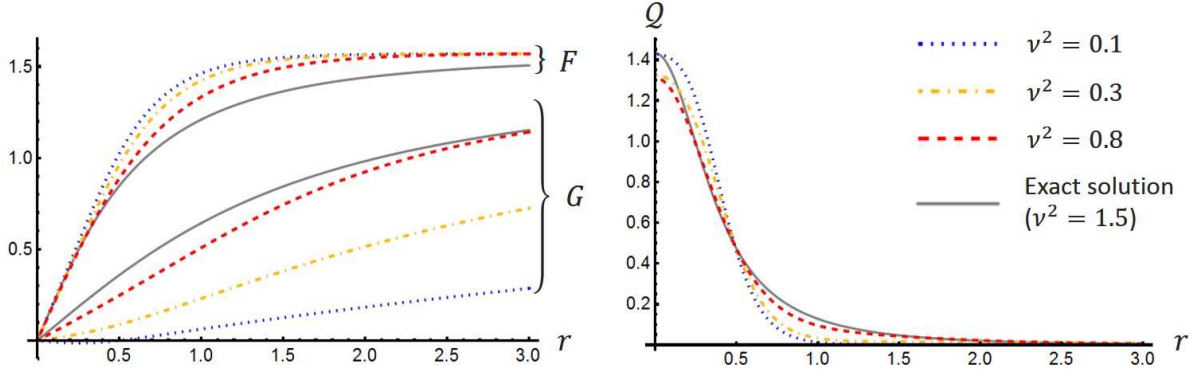}
		\caption{Plot of the profile functions $\{F, G\}$ (left) and the topological charge density (right) of numerical solutions for changing the coupling constant $\nu^2$ at $\mu^2=1.5$. The gray line indicates the quantities of the exact solution \eqref{sol_nonBPS_holomorphic} on the solvable line.  \label{Fig_profile} }
	\end{figure}

	\begin{table}[t!]
		\centering
		\begin{tabular}{|c|c|c|c|c|c|c|c|c|}
			\hline
			$\nu^2$ & $H$ & $H_{\rm D}$ & $H_{\rm L}$ & $\nu^2 H_{\rm ani}$ & $\mu^2 H_{\rm pot}$ & $H_{\rm Boundary}$ & Derrick & $Q$ \\ \hline
			0.1  & -117.47  & 13.51  & -136.48  & 125.49  & 5.67  & -125.67  & -2.00  & 2.00  \\ \hline
			0.3  & -34.02  & 13.41  & -53.60  & 41.37  & 6.69  & -41.89  & -1.99   & 2.00  \\ \hline
			0.8  & -8.46  & 13.06  & -29.37  & 14.73  & 8.82  & -15.71  & -1.91  & 2.00  \\ \bhline{1pt}
			1.5  & -4.19  & 12.57  & -16.76  & 1.09  & 15.66  & -16.76  & -2  & 2  \\ \hline
		\end{tabular}
		\caption{The Hamiltonian and topological charge for the numerical solutions with $\mu^2=1.5$ where "Derrick" denotes the value 
		$(H_{\rm L}+H_{\rm Boundary})/(\nu^2H_{\rm ani}+\mu^2H_{\rm pot})$, which is expected to be $-2$ by the scaling argument.
		For $\nu^2=1.5$, we used the exact solution \eqref{sol_nonBPS_holomorphic} so that the "Derrick" and topological charge for $\nu^2=1.5$ are exact values.}
	\end{table}	
	
	Let us consider the asymptotic behavior of the solutions of the equations~\eqref{equation_profile}.
	Near the origin, the leading terms in the power series expansion are
	\begin{equation}
		F\approx c_F~ r, \qquad G\approx c_G~ r, 
	\end{equation}
	where $c_F$ and $c_G$ are some constants implicitly depending on the coupling constants of the model.
	To see the behavior of solutions at large $r$, 
	we shift the profile functions as 
	\begin{equation}
	F = \frac{\pi}{2} - {\cal F},\qquad
	G = \frac{\pi}{2} - {\cal G}.
	\end{equation}
	Then, one 
	obtains 
	linearized asymptotic equations on the functions ${\cal F}$ and ${\cal G}$ of the forms
	\begin{equation}
	\begin{split}
	&\({\cal F}''+\frac{{\cal F}'}{r}-\frac{4{\cal F}}{r^2} \)
	+2\sqrt{2}\({\cal G}'-\frac{{\cal G}}{r} \) -2\( \nu^2 +3\mu^2\){\cal F} =0 ~,
	\\
	&\({\cal G}''+\frac{{\cal G}'}{r}-\frac{{\cal G}}{r^2} \)
	-2\sqrt{2}\({\cal F}'+\frac{2{\cal F}}{r} \)=0 ~.
	\end{split}
	\label{eq_linearlized}
	\end{equation}
	Unfortunately, the 
	equations \eqref{eq_linearlized} 
	may not support an analytical solution. However,
	these equations imply that the asymptotic behavior of the profile functions is similar to that of the functions \eqref{sol_nonBPS_holomorphic}, by a  replacement $\nu^2\kappa$ with $(\nu^2+3\mu^2)/4$. Indeed, the asymptotic equations \eqref{eq_linearlized} 
	depend on such a combination of the coupling constants, and there may exist an exact solution on the solvable line with the same  character of asymptotic decay as the localized soliton solution of the equation \eqref{equation_profile}.
	
	To implement a numerical integration of the coupled system of 
	ordinary differential equations~\eqref{equation_profile}, we introduce the normalized compact coordinate $X\in(0,1]$ via 
	\begin{equation}
	r = \frac{1-X}{X}.
	\end{equation} 
	The integration was performed by the Newton-Raphson method with the mesh point $N_{\rm MESH}=2000$.
	
	In 
	Fig.~\ref{Fig_profile}, we display some set of  numerical solutions for different values of the coupling  $\nu^2$ at $\mu^2 = 1.5$ and their topological charge density ${\cal Q}$ defined through $Q=2\pi\int r{\cal Q} \dd r$.
	The solutions enjoy Derrick's scaling relation and possess a good approximated value of the topological charge, as 
	shown in Table~1.
	One observes that as the value of the coupling $\nu^2$ becomes relatively small, the function $G$ is delocalizing while the profile function $F$ is approaching its vacuum value everywhere in space 
	except for the origin.
	This is an indication that any regular non-trivial solution does not exist $\nu^2=0$.

	\subsection{Asymptotic behavior}  
	
	Asymptotic interaction of solitons is related 
	to the overlapping of the tails of the profile functions of well-separated single solitons \cite{Manton:2004tk}. 
	Bounded multi-soliton configurations may exist if there is an attractive force between two isolated solitons.
	
	Considering the above-mentioned soliton solutions of 
	the gauged $CP^2$ NL$\sigma$-model, 
	 we  have seen that the exact solution 
	\eqref{sol_nonBPS_holomorphic} has the same type of asymptotic decay as any solution of the general system  \eqref{equation_profile}.
	Therefore, it is enough to examine the asymptotic force between the solutions on the solvable line \eqref{sol_nonBPS_holomorphic} to understand whether or not the Hamiltonian \eqref{Hamiltonian_generaln} supports multi-soliton solutions of higher topological degrees. Thus,  without loss of generality, we can set $\mu^2=\nu^2$.
	
	Following the approach discussed in Ref.~\cite{Manton:2004tk}, let us consider a superposition of two exact solutions above.
	This superposition is no longer a solution of the Euler-Lagrange equation, except for in the limit of infinite separation, because there is a force acting on the solitons. The interaction energy of two solitons  can be written as
	\begin{equation}
	E_{\rm int}(R) = H_{\rm sp}(R) - 2H_{\rm exact},
	\end{equation}
	where $H_{\rm sp}(R)$ is the energy of two BPS solitons separated by some large but finite distance $R$ from each other, and $H_{\rm exact}$ stands for the static energy of a single exact solution. 
	Notice that the lower bound of the  Hamiltonian \eqref{Hamiltonian_generaln} with $\mu^2=\nu^2$ is given 
	\begin{equation}
	H=\nu^{-2} H_{\nu^2=\mu^2=1} + (1-\nu^2) H_{\rm D} \geq 2\pi(1-2\nu^{-2})Q, 
	\end{equation}
	where the equality is enjoyed only by holomorphic solutions. 
	Therefore, we immediately conclude 
	\begin{equation}
	H_{\rm sp}(R)\geq 2H_{\rm exact},
	\end{equation} 
	where the equality is satisfied only at the limit $R\to\infty$.
	It follows that the interaction energy is always positive for finite separation, and the interaction is repulsive. 
	Since the exact solution has the topological charge $Q=2$, it implies that there are no isolated soliton solutions with the topological charge $Q\ge 4$ in this model.
	Note that, however, as the BPS solution \eqref{sol_BPS} suggests, there can exist soliton solutions with an arbitrary negative charge, which are topological excited states on top of the homogeneous vacuum state. 
	
	\section{Conclusion}
	\label{sec:conclusion}
	\setcounter{equation}{0}
	
	In this paper, we have studied two-dimensional Skyrmions in the $CP^2$ NL$\sigma$-model with a Lifshitz invariant term which is an $SU(3)$ generalization of the DM term.
	We have shown that the $SU(3)$ tilted FM Heisenberg model turns out to be an $SU(3)$ gauged $CP^2$ NL$\sigma$-model in which the term linear in a background gauge field can be viewed as a Lifshitz invariant. 
	We have found exact BPS-type solutions of the gauged $CP^2$ model in the presence of a potential term with a specific value of the coupling constant. The least energy configuration among the BPS solutions 
	has been discussed.
	We have reduced the gauged $CP^2$ model to the (ungauged) $CP^2$ model with a Lifshitz invariant by choosing a background gauge field.
	In the reduced model, we have constructed an exact solution for a special combination of coupling constants called the solvable line and numerical solutions for a wider range of them.
	
	For numerical study, we chose the background field, generating a potential term that can be interpreted as the quadratic Zeeman term or uniaxial anisotropic term.
	One can also choose a background field generating the Zeeman term; if the background field is chosen as $A_1=-\kappa\lambda_7$ and $A_2=\kappa\lambda_5$, the associated potential term is proportional to $\exv{S^3}$. 
	The Euler-Lagrange equation for the extended $CP^2$ model with this background field is not compatible with the axial symmetric ansatz \eqref{Ansatz_axial_sym}.
	Therefore, a two-dimensional full simulation is required to obtain a 
	solution with this background field. 
	This problem, numerical simulation for non-axial symmetric solutions in the $CP^2$ model with a Lifshitz invariant, is left to future study.
	In addition, the construction of a $CP^2$ Skyrmion lattice is a challenging problem. 
	The physical interpretation of the Lifshitz invariants is also an important future task.
	The microscopic derivation of the  $SU(3)$ tilted Heisenberg model \cite{zhu2014spin} may enable us to understand the physical interpretation and physical situation where the Lifshitz invariant appears.
	Other future work would be the extension of the present study to the $SU(3)$ antiferromagnetic Heisenberg model where soliton/sphaleron solutions can be constructed~\cite{Bykov:2015pka,Ueda2016, Amari2018}.

	We restricted our analysis on the case that the additional potential term $\mu^2H_{\rm pot}$ is balanced or dominant against the anisotropic potential term $\nu^2H_{\rm ani}$, i.e., $\nu^2 \leq \mu^2$. 
    We expect that a classical phase transition occurs outside of the condition, and it causes instability of the solution. At the moment, the phase structure of the model \eqref{Hamiltonian_generaln} is not clear, and we will discuss it in our subsequent work.

    Moreover, it has been reported that in some limit of a three-component Ginzburg-Landau model \cite{Garaud:2011zk,Garaud:2012pn}, and of a three-component Gross-Pitaevskii model \cite{Eto:2012rc,Eto:2013spa}, their vortex solutions can be well-described by planar $CP^2$ Skyrmions.
	We believe that our result provides a hint to introduce a Lifshitz invariant to the models, and that our solutions find applications not only in $SU(3)$ spin systems but also in superconductors and Bose-Einstein condensates described by the extended models, including the Lifshitz invariant.

	\vspace{2cm} 
 
    \textbf{Acknowledgments}\\
    This work was supported by JSPS KAKENHI Grant Nos. JP17K14352, JP20K14411, and JSPS Grant-in-Aid for Scientific Research on Innovative Areas ``Quantum Liquid Crystals'' (KAKENHI Grant No. JP20H05154). Ya.S. gratefully acknowledges support  by the Ministry of Education  of Russian Federation, project  FEWF-2020-0003.
	Y. Amari would like to thank Tokyo University of Science for its kind hospitality.
	
	\bibliographystyle{custom}
	\nocite{Schroers:2019hhe,Barton-Singer:2018dlh}
	\bibliography{reference}
	
\end{document}